\documentstyle[emulateapj, psfig]{article}
\begin{document}
\title{PG~1254+047: An Intrinsically X-ray Weak, Heavily Absorbed BALQSO?}
\author{Bassem M. Sabra \& Fred Hamann}
\affil{\textit{Department of Astronomy, University of Florida, Gainesville, 
FL 32611}}
\submitted{Accepted for publication in the Astrophysical Journal}

\begin{abstract}

We present \textit{Chandra} observations of the radio-quiet Broad Absorption 
Line (BAL) QSO PG~1254+047. We find that it is a weak X-ray
source, with a total of $44\pm 7$ photons measured in 36 ksec across
the observed energy range $0.7 - 6.0$~keV. Its X-ray weakness is consistent 
with the known correlation between $\alpha_{\rm ox}$ and the strength of the 
UV absorption lines. The spectral energy distribution suggests that 
PG~1254+047 is intrinsically X-ray weak, in addition to 
being heavily X-ray absorbed. The X-ray absorption column density is  
$N_{\rm H} > 10^{23}$ cm$^{-2}$ for neutral gas, while the intrinsic 
(unabsorbed) emission spectrum has $\alpha_{ox} > 2$. The data are fit best 
by including an ionized (rather than neutral) absorber, with column density  
$N_{\rm H} \gtrsim 2\times 10^{23}$ cm$^{-2}$. The degree of 
ionization is consistent with the UV BALs, as is the total column 
density if the strongest UV lines are saturated. If the X-ray absorber 
forms in a wind that is radiatively accelerated to the BAL velocities, then 
the wind must be 
launched from a radius of about $10^{16}$~cm with a mass loss rate of 
$\sim 1$~M$_\odot$~yr$^{-1}$.  
\end{abstract}
\keywords{galaxies: active---quasars: absorption lines---quasars: 
individual (PG~1254+047)---X-rays: galaxies}

\section{Introduction}

Almost all quasi-stellar objects (QSOs) are X-ray sources. The reprocessing 
of the X-rays by intervening matter along the line-of-sight to the QSO 
imprints informative features on the resulting spectrum (e.g., Turner 1991). 
X-ray absorption studies provide an important tool to determine the physical 
and chemical state of the gas associated with QSOs (e.g., Netzer 1996). 
A subclass of QSOs that have interesting X-ray properties in terms of 
continuum strength and absorption are the Broad Absorption Line (BAL) QSOs 
(Foltz et al. 1990; Weymann et al. 1991). They display broad 
(FWHM $\approx 10,000$~km~s$^{-1}$) absorption lines in the rest-frame 
ultraviolet (UV), which originate in an outflow of matter 
from the central engine of the QSO. The outflow velocity may reach up to 
$0.2\ c$ (Foltz et al. 1983). Determining the relation between 
the gases producing the X-ray and UV absorption has profound implications for  
the wind properties, such as geometry, launch radius, acceleration mechanism  
and mass loss rate (e.g., Mathur, Elvis, \& Singh 1995; Murray, Chiang, \& 
Grossman 1995; Hamann 1998). 

BALQSOs are also usually X-ray faint, especially in soft X-rays 
(Brandt, Laor, \& Wills 2000). The 
similarity of BALQSOs and non-BALQSOs in other parts of the spectrum 
suggest that X-ray weakness is due to heavy absorption. The high ionization 
state inferred from the UV absorption lines indicates that the X-ray 
absorbers are not generally neutral; they could be 
similar to the ``warm'' absorbers measured frequently in 
Seyfert 1 galaxies (George et al. 1998; Reynolds 1997). 

One major obstacle to understanding the relationship between
BALs and the X-ray absorbers has been that the total column densities
derived from the BALs (typically $N_H \lesssim 10^{20}$ cm$^{-2}$, Hamann,
Korista \& Morris 1993), are 3 or more orders of magnitude less
than those deduced from X-rays (cf. Brandt et al. 2000; Gallagher
et al. 1999; Green et al. 2001). However, there is now growing
evidence that the BALs are much more optically thick than previously
realized (Hamann 1998, Arav et al. 1999, Wang et al. 1999), and
therefore the total column densities in {\it outflowing} BAL gas
may be as large as the X-ray absorbers. The correlated appearance of
UV and X-ray absorption features (Brandt et al. 2000) shows clearly
that the two absorbers are (somehow) physically related. A key question
now is whether these features are manifestations of the same gas.

In this paper, we discuss Chandra X-ray observations of PG 1254+047. 
This object shows BALs ($z_{abs}=0.870$) that are blueshifted with respect to 
the systemic velocity of the QSO 
($z_{em}=1.024$). It is also characterized by being an X-ray faint source 
(Wilkes et al. 1994).
Hamann (1998) used the relative strengths of the BALs in
this source to show that 1) most of the lines are saturated and,
2) the total column density is {\it at least} $N_{\rm H} \approx 10^{22}$
cm$^{-2}$. Those results probably apply generally to BALQSOs, but
PG~1254+047 is of particular interest because its BALs are ``detached''
from the emission lines --- appearing at blueshifted velocities from
$\sim$15,000 to $\sim$27,000 km s$^{-1}$. If there is a high column
density X-ray absorber in this source that can be identified with
the BAL gas, then that X-ray absorber must also be outflowing at
minimally $15,000$ km s$^{-1}$. Such a finding would substantially impact
the derived mass loss rates and place severe constraints on models of
the wind acceleration (Hamann 1998; Hamann \& Brandt 2001). Our aim
here is to determine the properties of the X-ray 
spectrum, search for signs of absorption, and define the relationship 
between the UV and X-ray absorbing gas.  

\section{Observations and Data Reduction}

PG 1254+047 was observed by \textit{Chandra} using the Advanced Imaging 
Spectrometer (ACIS) on 29 May 2000. The most recent re-processed 
(29 February 2001) data released 
by the Chandra X-ray Center were used. No filtering for high background 
or bad aspect times was done because the light curves did not 
show any flare-ups in the count rate and the aspect was a smooth dither 
pattern with no outlying points. 
Data extraction and calibration was done using version 
1.4 of the Chandra Interactive Analysis of Observations (CIAO). 
The X-ray analysis package XSPEC (Arnaud 1996) 
was used for rebinning and spectral analysis. 
We created the response matrix and ancillary response files by relying 
on calibration data provided when the chip temperature during observations 
was $-120^{\circ} $ C.

We extracted the source counts from a circular region of radius of 
$5\arcsec$ while the 
background region was an annulus with radii between $10\arcsec$ and 
$20\arcsec$, both centered on the position of PG~1254+047. 
The ACIS image is shown in Figure 1. As determined from the 
ACIS image, the position of PG~1254+047 is $\sim$1\arcsec --- 
comparable to the spatial resolution of ACIS --- from the optical 
position of the source 
($\alpha({\rm J}2000)=12^h 56^m 59.96^s, 
\delta({\rm J}2000)=04^\circ 27\arcmin 34.2\arcsec$, 
Gould, Bahcall, Maoz 1993). 
The mere fact of actually detecting 
PG~1254+047 in X-rays is notable because BALQSOs are expected to be 
extremely weak in X-rays (Brandt et al. 2000). 
We obtained a total of $44\pm 7$ counts in an effective exposure time of 
36 ksec across the observed energy range $\sim 0.7-6.0$~keV. 

\section{Analysis and Results}

We want to understand the reason behind the low count rate: 
Is it due to intrinsic absorption, intrinsic X-ray weakness, or both? 
Given the paucity of photons detected, our approach is that of 
elimination, starting with the simplest model and increasing 
the level of complexity only after exhausting all means to satisfactorily 
fit the data with a given model. Throughout the analysis we assume 
solar abundances. We also use XSPEC for $\chi^2$ minimization after 
binning the spectrum to have at least 10 counts/bin. Based on 
experimentation, we find that $\chi^2$ analysis leads to outcomes 
similar to Cash statistics.  The results of $\chi^2$ statistics are 
listed in Table 1. We explain below the models in more detail. 

{\it Model A}: We fit the data with a power law continuum 
of the form $\phi_E = A~(\frac{E}{1~{\rm keV}})^{-\Gamma}$, 
where $A$ is the normalization of the power law at 1~keV, in 
units of photons cm$^{-2}$ s$^{-1}$ keV$^{-1}$, and $\Gamma$ is the X-ray 
photon index, absorbed by a Galactic column density of 
$N^{Gal.}_{\rm H}=2\times 10^{20}$ cm$^{-2}$ (Lockman \& Savage 1995).  
Both the normalization and the photon index were left as free 
parameters (Figure 2). The slope was found to be rather flat, 
$\Gamma = 0.36$, for a QSO, where usually 
$1.3 < \Gamma < 2.3$ in the observed frame of 0.5-10 keV 
(Reeves et al. 1997). 
The flatness of the X-ray slope suggests that there is 
additional intrinsic absorption. 
 
{\it Model B}: We adopt a ``normal'' QSO continuum, 
specified by  $\Gamma=1.9$ and $\alpha_{\rm ox}=1.6$ (Laor et al. 1997), 
with $\alpha_{ox}=0.384 \log(f_\nu(2500\ {\rm \AA})/f_\nu (2\ {\rm keV}))$, 
where $f_\nu(2500\ {\rm \AA})$ and $f_\nu (2\ {\rm keV})$ are the rest-frame 
specific fluxes at 2500~\AA\ and 2~keV, respectively. We attenuate this 
continuum  through an intrinsic  neutral absorber, with $N^{int}_{\rm H}$, 
at $z_{em}$. We derive the 
rest-frame $f_\nu(2500\ {\rm \AA})$ from its B-magnitude, 
including the appropriate Galactic dereddening, 
$E(B-V)=N^{Gal.}_{\rm H}/5.27\times 10^{21}=0.09$ (Lockman \& Savage 1995), 
and the k-correction (see Green 1996). 
The choice of $\alpha_{ox}$ then determines the 
normalization of the power law at the observed energy of 2~keV$/(1+z)$. 
In Figure 3, the poor fit at low energies suggests intrinsic absorption, 
while the bad overall fit indicates that the observed X-ray weakness 
cannot be explained by absorption alone.  

The bad overall fit indicates that PG~1254+047 is both 
intrinsically X-ray weak and heavily absorbed. The large discrepancy 
between the data and model is due to the fact that absorption not 
only decreases the X-ray flux but also deforms the intrinsic spectrum, 
where the effect is most pronounced at low energies. 

{\it Model C}: To study the possibility of both intrinsic X-ray weakness and 
absorption, we remove the constraint that $\alpha_{\rm ox}=1.6$ and hence 
allow the normalization of the power law to vary. The fits improve 
drastically, though not to the extent of giving an acceptable fit 
($\chi_\nu^2 \approx 1$). We show the results in Figure 4. It is 
evident that the fit does not strongly support neutral 
absorption due to the large discrepancy between the data and the models at 
soft energies. Also, we know that there is no neutral absorber with the 
above quoted column density because the UV spectra do not contain 
low-ionization metal lines (Hamann 1998). 

{\it Model D}: We explore absorption by ionized gas. 
We hereafter leave the normalization as a free parameter. The amount  
of absorption depends on the intrinsic total hydrogen column density 
and the ionization parameter, $U$, defined as the ratio of the density of 
hydrogen ionizing photons to that of hydrogen particles (H$^0$ + H$^+$). 
We model the 
ionized absorber using the photoionization code CLOUDY (Ferland et al. 1998). 
We generate a grid of QSO continua attenuated through ionized absorbers 
described by a grid of $U$ and $N_{\rm H}$. The incident continuum is a 
piecewise powerlaw (Zheng et al. 1996; Laor et al. 1997), which is 
similar to that used in Hamann (1998). The ultimate results are not 
critically dependent on the shape the continuum, as long as it 
in general agreement with that of QSOs (Hamann 1997). 
We then use XSPEC to interpolate 
on this grid to find the best fitting model of the observed continuum. 
Hamann (1998) derived the range of $U$ and $N_{\rm H}$ values that are 
consistent with the measured BALs (his Figure 7). Larger total $N_{\rm H}$ 
must be accompanied by large $U$. 
For Model D we fix $\log U=1.2$ to be consistent with both 
the BAL data and the large X-ray column densities derived here. 
The resulting fit (Figure 5) improves considerably 
with $\chi_\nu^2 = 0.20$ and $N_{\rm H}=2.8\times 10^{23}$~cm$^{-2}$.  
The value of $A$ 
corresponds to an intrinsic (unabsorbed) $\alpha_{\rm ox}=2.2$, suggesting 
that PG~1254+047 is intrinsically X-ray weak. We comment more on this in the 
following section. 

{\it Model E}: It is similar to Model D, except that we use a 
steeper powerlaw index of $\Gamma=2.5$. The final results do not 
deviate a lot from those of Model D. We experimented with this 
steep spectrum to see if we can evade the requirement of intrinsic 
X-ray weakness which the previous models suggested (see discussion below). 

For the models with intrinsic absorption (B-E), we also experimented with 
putting 
the X-ray absorber at the redshift of the UV absorption lines instead of at 
the systemic velocity of the QSO. This redshift difference is unresolvable 
with ACIS. Nonetheless placing the X-ray absorber at $z_{abs}$ leads 
to lower column densities (by a factor of less than 2). 

\section{Discussion}
\subsection{Intrinsic X-ray Weakness}
The results presented in the previous section require further discussion. 
They suggest that PG~1254+047 is both intrinsically 
X-ray weak --- intrinsic $\alpha_{\rm ox}=2.2\pm 0.1$ obtained after 
correcting for absorption --- and heavily absorbed. 
Recent observations of other BALQSOs 
indicate that the X-ray weakness is due to absorption by large column 
densities (e.g., Brandt et al. 2000; Green et al. 2001; Mathur et al. 2001). 
In particular, Brandt et al. (2000) discovered an anti-correlation between 
the the rest-frame equivalent width of \ion{C}{4}$\lambda\lambda 1548,1550$ 
and $\alpha_{\rm ox}$; strong absorption lines are usually accompanied by 
an X-ray absorber that steepens 
$\alpha_{ox}$. For PG~1254+047, $\alpha_{\rm ox} = 2.6 \pm 0.1$ 
(uncorrected for intrinsic absorption, Model A in Table 1)  and the 
equivalent width of \ion{C}{4} is $\sim 13$~\AA\ (Hamann 1998). 
These values are consistent with the anticorrelation in Brandt et al. (2000).  

However, the observed, absorption corrected flux from PG~1254+047,  
$F(0.2-10~{\rm keV})=5.2\times 10^{-14}$~erg~s$^{-1}$~cm$^{-2}$, 
is about a factor of 29 below that predicted on the basis of its  
B-magnitude and $\alpha_{\rm ox} = 1.6$. We have shown in the previous 
section that no amount of intrinsic 
absorption was capable of decreasing this amount of intrinsic flux, 
while at the same time reproducing the overall X-ray spectral shape. 

One possible way to decrease the observed X-ray flux without changing 
$\alpha_{ox}$ is if the X-ray spectrum declines sharply towards higher 
energies. Mathur et al. (2001) presented a 
deep ASCA spectrum of the BALQSO PHL 5200, one of the X-ray brightest of its 
class, and concluded that the 
X-ray weakness is due to heavy absorption 
($N_{\rm H}=5\times 10^{23}$~cm$^{-2}$) 
and that the X-ray spectrum is very steep ($\Gamma=2.5$). We experimented 
with various values of $\Gamma$ ranging from 2.5 to 10 while freezing 
$\alpha_{\rm ox}$ at 1.6, together with intrinsic absorption (neutral and 
ionized). None of the fits was found to be acceptable; $\chi^2_\nu$ 
did not decrease to below $\sim 3.5$. We note that $\Gamma=2.5$ and 
$\alpha_{\rm ox}=2.0$ lead to satisfactory results 
($\chi^2_\nu \approx 0.63$, Model E). Partially covering neutral/ionized 
absorption (Hamann 1998) also did not lead to significant improvement.  
 
In principle, there is no physical reason that precludes the occurence 
of intrinsically weak X-ray AGNs. This weakness could be due to time 
variability, i.e. a QSO caught at an X-ray minimum. For example, Narrow Line 
Seyfert 1s (NLSy1s)
show rapid variability and steep soft X-ray spectra.  Mathur (2000) draws the 
analogy between  BALQSOs and NLSy1s arguing that BALQSOs are AGNs in their 
early evolutionary states, accreting at close to their Eddington limits. 
Leighly et al. (2001) report a {\it Beppo}Sax observation of a 
very bright ($B=13.9$) QSO, identified from the FIRST survey. PHL 1811 
has optical characteristics similar to those of NLSy1s and is extremely X-ray 
weak with a steep powerlaw slope ($\Gamma = 2.6$). PG~1254+047 could have been 
observed while it was in a quiescent X-ray state. 

X-ray weakness could also be due to the smallness of the  
X-ray emitting region (Mathur 2000; Leighly et al. 2001).  Another 
possibility for X-ray faintness is that we are actually not 
seeing all the X-rays coming from the source, but instead the observed X-rays 
are the small fraction that have been scattered into our line of sight by an 
electron mirror (Gallagher et al. 1999). To agree with our data, the direct 
X-rays from the nucleus must be completely absorbed by a Thomson thick 
absorber, 
while the scattered X-rays are the ones being attenuated through the 
ionized absorber. The UV-absorbing gas must be outside of the Thomson thick 
medium. The relation between the UV and X-ray absorbing gases is discussed 
below. 

\subsection{Wind Dynamics}
Putting the issue of X-ray weakness aside, we can still make important 
conclusions about the UV/X-ray absorber connection and the physics of wind 
generation. It is natural to suppose that the X-ray absorption occurs in the 
outflowing BAL gas (Mathur et al. 1995). However, standard analysis of the 
BALs, which uses the absorption line troughs to derive optical depths and 
column densities, typically implies total columns in the range 
$10^{19} \lesssim N_{\rm H} \lesssim 10^{20}$~cm$^{-2}$ -- two or more orders 
of magnitude less than the X-ray absorbing columns (Hamann, Korista, \& 
Morris 1993 and references therein)! This enormous 
discrepancy between the UV and X-ray data must be resolved if we are to reach 
even a rudimentary understanding of BAL outflows. 

All these problems can be addressed by a direct comparison between the 
UV and X-ray data.
Hamann (1998) presented a new analysis of 
\ion{P}{5}$\lambda\lambda 1118, 1128$ and other BALs in 
high-quality HST spectra of PG~1254+047. He used explicit 
calculations of the column densities and line optical depths under different 
ionization conditions to show that, if the metals have roughly solar  
relative abundances, then the significant presence of PV absorption implies 
that 1) many of the strong lines like \ion{C}{4}, \ion{N}{5} and \ion{O}{6}  
are much more optically thick than they appear, and 2) 
the total column density in the BAL gas is 
$N_{\rm H}\gtrsim 10^{22}$~cm$^{-2}$, considerably larger than previously 
expected from the depths of the measured troughs. 
The large column densities and line optical depths are disguised in the 
observed (moderate-strength) BALs because the absorber does not fully cover 
the background light source(s) along our line(s) of sight. 
Our choice of $U$ in Model D leads to an $N_{\rm H}$ that is in agreement with 
the UV data (Hamann 1998). The UV and X-ray absorbers might,
therefore, be identical. 

If the UV and X-ray absorption is arising from the same gas, then the total 
hydrogen column density deduced from the X-rays places additional 
constraints on the acceleration mechanism of the BAL  
Region. A relation exists between the terminal velocity of a 
radiatively accelerated outflow, its column density, and the radius 
from which it was launched (cf. equation 3 of Hamann 1998). Large column 
densities require a small launch radius for sufficient radiative 
acceleration. Substituting values appropriate to PG~1254+047 
($M_{\rm BH}= 10^8$~M$_\odot$, $L= 10^{46}$~ergs s$^{-1}$ cm$^{-2}$, 
$v_{terminal}=20,000$ km s$^{-1}$, and $N_{\rm H}=2.8 \times 10^{23}$ 
cm$^{-2}$), we find that the inner radius is $\sim 10^{16}$~cm. 
Therefore, the wind is launched from very close to the blackhole. We also 
calculate the mass loss rate to be $\sim 1$~M$_\odot$~yr$^{-1}$. We use 
a relation presented in Hamann \& Brandt (2001) to derive the hydrogen 
volume density given $U$, a radius, and $L_{46}$. Taking our 
above values for $U$ (Model D) and $L_{46}$, and assuming that the 
BALs form at twice the launch radius, we find that 
$n_{\rm H}\approx 5\times 10^{10}$ cm$^{-3}$, which yields a BAL region 
thickness of at least $6\times 10^{12}$ cm, using $N_{\rm H}$ from Model D, 
in agreement with parameters derived for other BALQSOs 
(Hamann \& Brandt 2001).   

A very strong constraint from the UV data is the fact that the BALs are 
detached (Hamann 1998), i.e. the troughs extend from $-15,000$ to 
$-27,000$~km~s$^{-1}$, with no absorption near the systemic rest velocity. 
The BALs thus imply that the outflowing gas has already been accelerated
to those tremendous speeds. The situation might be like that described 
Murray et al. (1995), Murray \& Chiang (1995) and Elvis (2000). 
The wind is launched vertically, perpendicular to 
the accretion disk, and then it bends and flares radially outwards 
as it is accelerated by radiation pressure from the central engine. 
The bent geometry provides a natural explaination for detached BAL 
troughs; lines of sight away from the disk plane can sample portions of
the wind that have already been accelerated to the high observed speeds.
In the Murray et al. (1995) model, an X-ray absorber, essentially at rest, 
shields the wind downstream from soft X-rays. This shielding allows the 
BAL wind to maintain a low enough ionization parameter for substantial
resonant line driving and acceleration up to speeds 
of $\sim 0.1~c$ (Murray et al. 1995). Alternatively, the wind may be 
capable of shielding itself (Murray \& Chiang 1995; Elvis 2000 ), in
which case the UV and X-ray absorbers are identical and the large column
densities we infer from the X-rays must be part of the high velocity
wind. The dynamical implication of these large {\it outflowing} column
densities is a small launch radius, as derived above and proposed  
originally by Murray et al. (1995).

\noindent {\it Acknowledgements:} We wish to acknowledge support 
through {\it Chandra} grants GO 0-1123X and GO 0-1157X. We thank Ian 
George for comments which improved the presentation, and Keith Arnaud, 
Niel Brandt, Norman Murray, and Beverley Wills for useful discussion.

\newpage
\begin{figure}
\vbox{
\centerline{
\psfig{figure=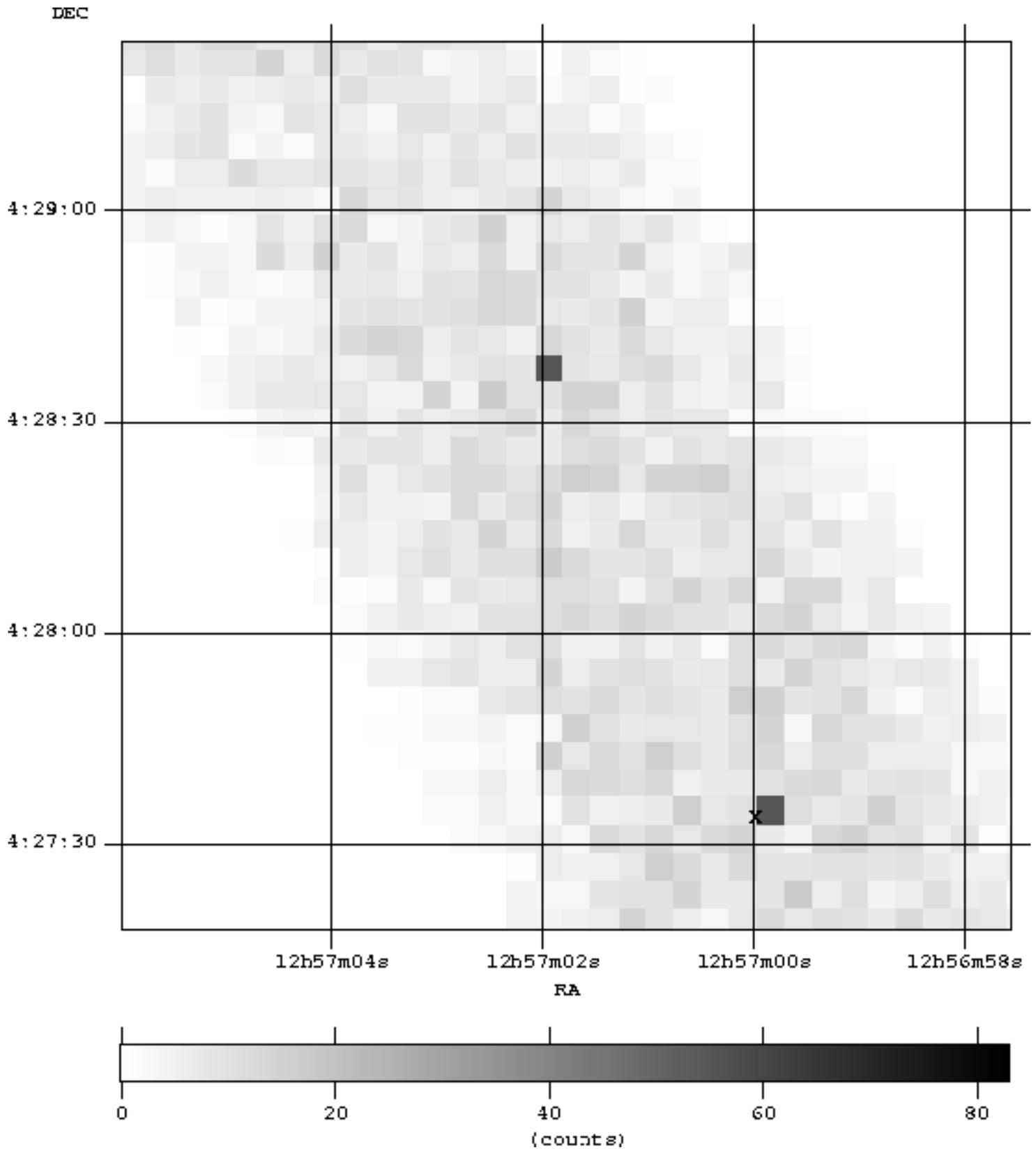}
}
\caption{ACIS image of PG 1254+047. PG~1254+047 is the darkest pixel nearest 
to $\alpha({\rm J}2000)=12^h 57^m 00^s$. 
The ``X'' symbol denotes the optical position of the 
QSO, within 1\arcsec of the X-ray position. The other dark pixel is 
$\sim 13\arcsec$ from a visual source reported in Kirhakos et al. (1994).} 
}
\end{figure}

\begin{figure}
\vbox{
\centerline{
\psfig{figure=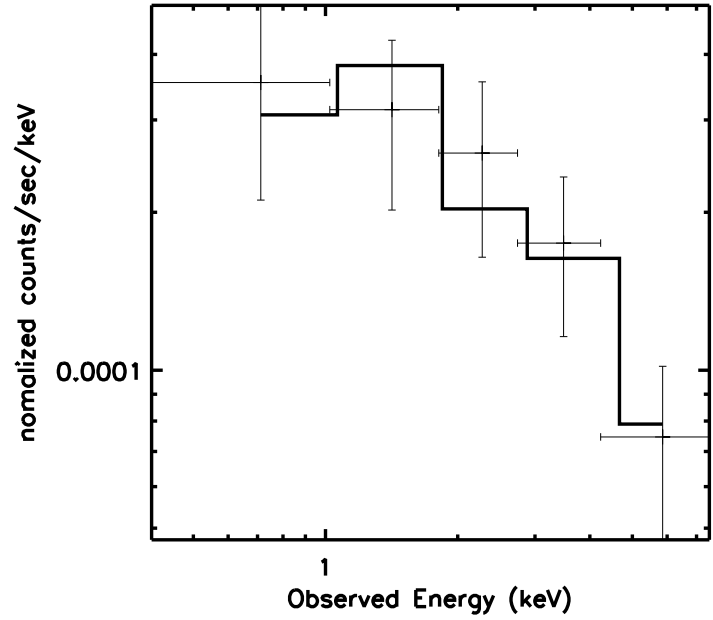,width=10cm}
}
\caption{Model A: Galactic Absorption only with free normalization. The 
crosses represent the data binned to provide 10 counts per bin. The solid 
histogram is the fit. The fit yields   
$A= 6.47\times 10^{-7}$ photon~s$^{-1}$~cm$^{-2}$~keV$^{-1}$ at 1~keV, 
and $\Gamma=0.36$,  for $\chi_\nu^2=0.29$ and 3 degrees of freedom (d.o.f.).}
}
\end{figure}

\begin{figure}
\vbox{
\centerline{
\psfig{figure=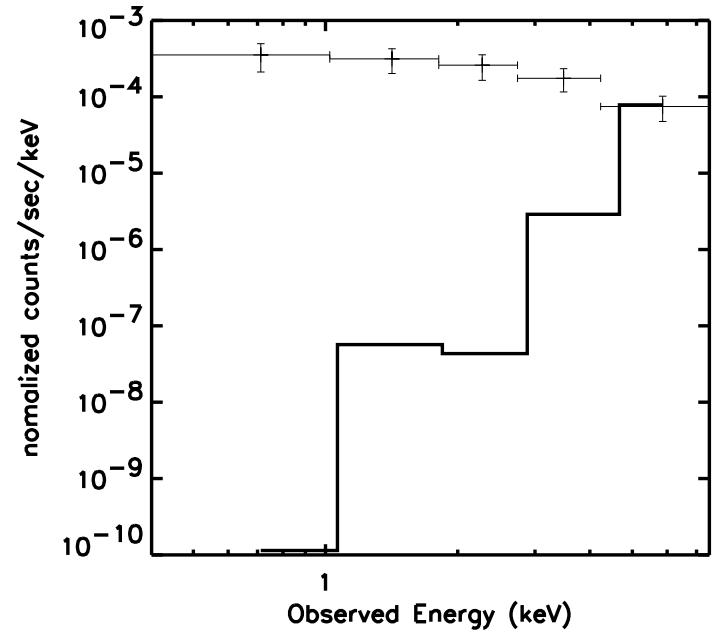,width=10cm}
}
\caption
{Model B: Neutral Absorber and fixed normalization.   
$N_H=5.01\times 10^{24}$cm$^{-2}$, $\chi_\nu^2=7.50$ for 4 d.o.f. }
}
\end{figure}

\begin{figure}
\vbox{
\centerline{
\psfig{figure=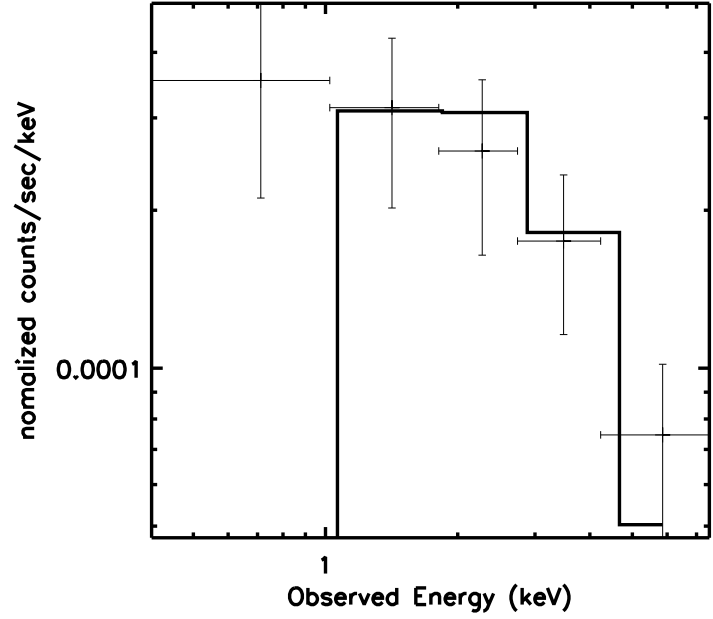,width=10cm}
}
\caption
{Model C: Neutral Absorber and free normalization.   
$N_H=10.81\times 10^{22}$cm$^{-2}$,  
$A = 5.86\times 10^{-6}$ photon~s$^{-1}$~cm$^{-2}$~keV$^{-1}$ at 1~keV, 
$\chi_\nu^2=2.28$ for 3 d.o.f.}
}
\end{figure}

\begin{figure}
\vbox{
\centerline{
\psfig{figure=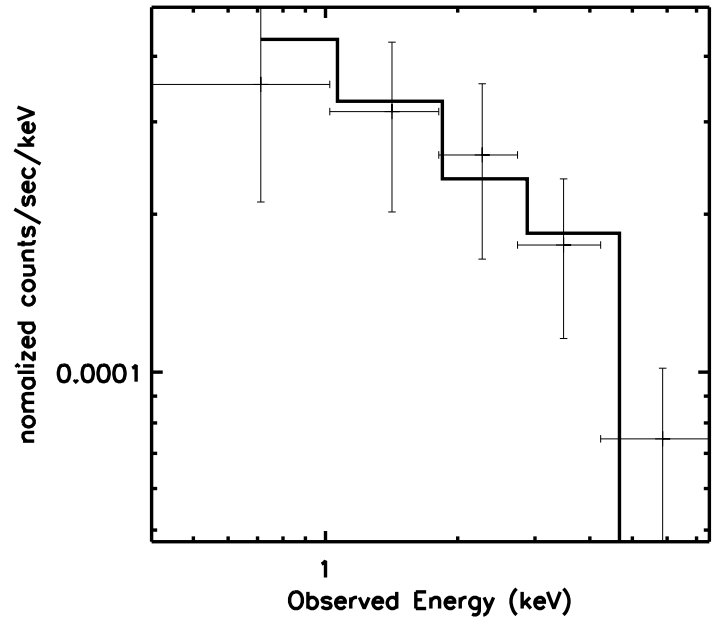,width=10cm}
}
\caption
{Model D: Ionized Absorber.   
$N_H=2.8\times 10^{23}$cm$^{-2}$, $\log U=1.2$,  
$A= 8\times 10^{-6}$ photon~s$^{-1}$~cm$^{-2}$~keV$^{-1}$ at 1~keV, 
$\chi_\nu^2=0.20$ for 3 d.o.f. }
}
\end{figure}

\begin{table}
\begin{center}
\caption{Spectral Fit Parameters$^a$}
\vspace*{0.1in}
\begin{tabular}{ccccccc}
\tableline
\tableline
Model & $\log U$ & $N_{\rm H}$ ($10^{22}$ cm$^{-2}$) & $A^b$  & $\Gamma$ &$\alpha_{ox}$ &$\chi_\nu^2$ \\
\tableline
A     & -----  &  -----  & $0.7^{+0.2}_{-0.3}$  & $0.36^{+0.42}_{-0.44}$ & $2.6\pm 0.1$ & 0.29 for 3 d.o.f$^e$\\
B     &neutral & $501^d$     & $240^c$  & $1.9^c$  & $1.6^c$ & 7.50 for 4 d.o.f\\
C     &neutral & $10.8^d$    & $5.9^d$  & $1.9^c$  & $2.2^d$ & 2.28 for 3 d.o.f\\
D     &$1.2^c$ & $28^{+3}_{-4}$     & $8\pm 3$  & $1.9^c$  & $2.2\pm 0.1$ & 0.20 for 3 d.o.f\\
E     &$1.2^c$ & $28^{+2}_{-4}$  & $17\pm 7$   & $2.5^c$ & $2.0\pm 0.1$ & 0.63 for 3 d.o.f\\
\tableline
\end{tabular}
\end{center}
$^a$ All errors are at the 90\% confidence level.\\
$^b$ In units of $10^{-6}$ photon~s$^{-1}$~cm$^{-2}$~keV$^{-1}$.\\
$^c$ Fixed parameter.\\
$^d$ Confidence levels not quoted since $\chi_\nu^2 > 2$.\\
$^e$ degrees of freedom.
\end{table}


\begin{references}
\reference{} Arav, N., Becker, R. H., Laurent-Muehleisen, S. A., 
Gregg, M. D., White, R. L., Brotherton, M. S., \& de Kook, M. 
1999, \apj, 524, 566
\reference{} Arnaud, K. A. 1996, Astronomical Data Analysis Software and 
Systems V, eds. Jacoby G. and Barnes J., p17, ASP Conf. Series volume 101
\reference{} Brandt, W. N., Laor, A., \& Wills, B. J. 2000, \apj, 528, 637
\reference{} Elvis, M. 2000, \apj, 545, 63
\reference{} Ferland, G., et al. 1998, \pasp, 110, 1040
\reference{} Foltz, C. B., Wilkes, B., Weymann, R., \& Turnshek, D. 1983, 
\pasp, 95, 341
\reference{} Foltz, C. B., Chaffee, F. H., Hewett, P. C., Weymann, R. J., \& 
Morris, S. L. 1990, \baas, 2, 806
\reference{} Gallagher, S. C., Brandt, W. N., Sambruna, R. M., Mathur, S., \& 
Yamasaki, N.  1999, \apj, 519, 549
\reference{} George, I. A., Turner, T. J., Netzer, H., Nandra, K., 
Mushotzky, R. F., \& Yaqoob, T. 1998, \apjs, 114, 73
\reference{} Gould, A., Bahcall, J. N., \& Maoz, D. 1993, \apjs, 88, 53
\reference{} Green, P. J., Aldcroft, T. L., Mathur, S., Wilkes, B. J, \& 
Elvis, M. 2001, \apj\ accepted, astro-ph/0105258
\reference{} Hamann, F., Korista, K. T., \& Morris, S. L. 1993, \apj, 415, 541
\reference{} Hamann, F. 1997, \apjs, 109, 279
\reference{} Hamann, F. 1998, \apj, 500, 798
\reference{} Hamann, F., \& Brandt, W. N. 2001, in prep.
\reference{} Kirhakos, S., et al. 1994, \pasp, 106, 646
\reference{} Laor, A., et al. 1997, \apj, 477, 93
\reference{} Leighly, K. M., Halpern, J. P., Helfand, D. J., Becker, R. H., 
\& Impey, C. D. 2001, \aj, in press
\reference{} Lockman, F. J., \& Savage, B. D. 1995, \apjs, 97, 1
\reference{} Mathur, S., Elvis, M., \& Singh, K. 1995, \apj, 455, L9
\reference{} Mathur, S. 2000, \mnras, 314, L17 
\reference{} Mathur, S., Matt, G., Green, P. J., Elvis, M., Singh, K. P. 2001, 
\apj, in press
\reference{} Murray, N., \& Chiang, J. 1995, \apj, 454, L101
\reference{} Murray, N., Chiang, J., Grossman, J. A., \& Voit, G. M. 1995, 
\apj, 451, L498
\reference{} Netzer, H. 1996, \apj, 473, 781
\reference{} Reeves, J. N., et al. 1997, \mnras, 292, 468
\reference{} Reynolds, C. S. 1997, \mnras, 286, 513
\reference{} Turner, T. J., Weaver, K. A., Mushotzky, R. F., Holt, S. S., 
\& Madejski, G. M., 1991, \apj, 381, 85 
\reference{} Wang, T. G., Wang, J. X., Brinkmann, 
\& Matsuoka, M. 1999, \apj, 519, L35
\reference{} Weymann, R. J., Morris, S. L., Foltz, C. B., \& Hewett,
 P. C. 1991, \apj, 373, 23
\reference{} Wilkes, B. J., Tananbaum, H., Worral, D. M., Avni, Y., Oey, 
M. S. \& Flanagan, J. 1994, \apjs, 92, 53
\reference{} Zheng, W., et al. 1996, \apj, 475, 469\\
\end{references}
\end{document}